\documentclass{revtex4}
\usepackage{amssymb}
\usepackage{amsmath}
\usepackage{graphicx}
\usepackage{amsmath}
\usepackage[colorlinks]{hyperref}

\begin{document}

\title{Bulk viscous FWR with time varying constants revisited}
\author{J.A. Belinch\'{o}n}
\email{abelcal@ciccp.es}
\affiliation{Dpt. of Physics ETS
Architecture. UPM Madrid. Av. Juan de Herrera N-4 28040 Espa\~na.}
\date{\today}

\begin{abstract}
We study a full causal bulk viscous cosmological model with flat FRW
symmetries and where the ``constants'' $G,c$ and $\Lambda $ vary. We take
into account the possible effects of a $c-$variable into the curvature
tensor in order to outline the field equations. Using the Lie method we find
the possible forms of the ``constants'' $G$ and $c$ that make integrable the
field equations as well as the equation of state for the viscous parameter.
It is found that $G,c$ and $\Lambda $ follow a power law solution verifying
the relationship $G/c^{2}=\kappa .$ Once these possible forms have been
obtained we calculate the thermodynamical quantities of the model in order
to determine the possible values of the parameters that govern the
quantities, finding that only a growing $G$ and $c$ are possible while $%
\Lambda $ behaves as a negative decreasing function.
\end{abstract}

\maketitle

\section{Introduction}

In a recent paper (see \cite{Tony1}) we study a cosmological model with flat
FRW symmetries filled by a perfect fluid and where the ``constants'' $G,c$
and $\Lambda $ were considerate as function on time $t$. By different
reasons exposed in such paper, we took into account the possible effects of
a $c-$variable into the curvature tensor in order to outline the field
equations. Through the Lie group method we studied the possible forms of the
functions $G$ and $c$ that make integrable the field equations. In this way
we were finding that $G$ and $c$ follow a power law solution verifying the
relationship $G/c^{2}=\kappa .$ But unfortunately we were not able to
determine if $G$ and $c$ are growing or decreasing functions on time $t.$

In order to determine if $G$ and $c$ are growing or decreasing functions we
try to take into account thermodynamical considerations, that is to say, we
hope that thermodynamical restrictions help us to determine the behaviour of
such functions. For this purpose, in this paper we consider a cosmological
model with flat FRW symmetries filled by a full causal bulk viscous fluid
and where the constants $G,c$ and $\Lambda $ are considered as functions on
time $t.$ Once we have outlined the field equations (taking into account the
possible effects of a $c-$variable into the curvature tensor) we rewrite
them in order to obtain a second order differential equation in order to
apply the standard Lie procedure. The study of this ode through the Lie
group method allows us to obtain the precise form of the functions $G$ and $%
c $ that make integrable the field equations as well as the equation of
state for the bulk viscous parameter $\xi .$

As we will see in section 3 the field equations only admit scaling
symmetries (note that we are working with flat FRW symmetries) i.e. we are
studying a self-similar model. This fact obligates that $G$ and $c$ follow a
power law solution, $c=c_{0}t^{K_{1}},$ verifying the relationship $%
G/c^{2}=\kappa ,$ i.e. $G$ and $c$ are functions on time $t$ but in such a
way that this relationship must be verified. We find another restriction
under this symmetry. The bulk viscosity $\xi ,$ must follow the law $\xi
=k_{\gamma }\rho ^{1/2}$, i.e. we have found a concrete equation of state
for the viscous parameter $\gamma =1/2.$ All these results are in agreement
with our previous paper (\cite{Tony2}).

Once we have found the possible forms of the functions $G$ and $c$ we
calculate the energy density finding in a first approach that the only
physical solution imply that $G$ and $c$ must be a growing functions on time
$t$ since $K_{1}>0.$

In section 4 we will calculate all the physical quantities in order to
complete the solution for our model. In particular we are interested in
calculating the entropy in order to obtain some restrictions for the
physical parameters as $K_{1}$ and to elucidate if $G$ and $c$ are growing
or decreasing functions. In this way we find an exact solution for $\Lambda $
which behaves as a negative decreasing function.

We end summarizing all these results in the last section.

\section{The Model\label{M}}

Following Maartens \cite{Ma95}, we consider a Friedmann-Robertson-Walker
(FRW) Universe with a line element
\begin{equation}
ds^{2}=c^{2}(t)dt^{2}-f^{2}(t)\left( dx^{2}+dy^{2}+dz^{2}\right) ,
\label{line}
\end{equation}
filled with a bulk viscous cosmological fluid with the following
energy-momentum tensor:
\begin{equation}
T_{i}^{k}=\left( \rho +p+\Pi \right) u_{i}u^{k}-\left( p+\Pi \right) \delta
_{i}^{k},  \label{1}
\end{equation}
where $\rho $ is the energy density, $p$ the thermodynamic pressure, $\Pi $
is the bulk viscous pressure and $u_{i}$ is the four velocity satisfying the
condition $u_{i}u^{i}=1$.

The gravitational field equations with variable $G$, $c$ and $\Lambda $ are:
\begin{equation}
R_{ik}-\frac{1}{2}g_{ik}R=\frac{8\pi G(t)}{c^{4}\left( t\right) }%
T_{ik}+\Lambda (t)g_{ik}.  \label{ECU1}
\end{equation}
Applying the covariance divergence to the second member of equation (\ref
{ECU1}) we get:
\begin{equation}
div\left( \frac{G}{c^{4}}T_{i}^{j}+\delta _{i}^{j}\Lambda \right) =0,
\label{conser1}
\end{equation}
\begin{equation}
T_{i;j}^{j}=\left( \frac{4c_{,j}}{c}-\frac{G_{,j}}{G}\right) T_{i}^{j}-\frac{%
c^{4}\delta _{i}^{j}\Lambda _{,j}}{8\pi G},  \label{conser2}
\end{equation}
that simplifies to:
\begin{equation}
\dot{\rho}+3\left( \rho +p\right) H+3H\Pi =-\frac{\overset{\cdot }{\Lambda }%
c^{4}}{8\pi G}-\rho \frac{\dot{G}}{G}-4\rho \frac{\dot{c}}{c},
\label{conser3}
\end{equation}
where $H$ stands for the Hubble parameter $({\mathit{H=\dot{f}/f}}).$
Therefore, our model (with FRW symmetries) is described by the following
equations:
\begin{align}
2\dot{H}-2\frac{\dot{c}}{c}H+3H^{2}& =-\frac{8\pi G}{c^{2}}\left( p+\Pi
\right) +\Lambda c^{2},  \label{field1} \\
3H^{2}& =\frac{8\pi G}{c^{2}}\rho +\Lambda c^{2},  \label{field2} \\
\dot{\rho}+3\left( \rho +p+\Pi \right) H& =-\frac{\dot{\Lambda}c^{4}}{8\pi G}%
-\rho \frac{\dot{G}}{G}+4\rho \frac{\dot{c}}{c},  \label{field3} \\
\tau \dot{\Pi}+\Pi & =-3\xi H-\frac{\epsilon }{2}\tau \Pi \left( 3H+\frac{%
\dot{\tau}}{\tau }-\frac{\dot{\xi}}{\xi }-\frac{\dot{T}}{T}\right) .
\label{field4}
\end{align}

We would like to emphasize that deriving (\ref{field2}) and taking into
account (\ref{field1}) it is obtained (\ref{field3}), that is to say, the
conservation equation (\ref{field3}) could be deduced from the field
equation as in the standard case where the ``constants'' $G,c$ and $\Lambda $
are true constants (see \cite{Tony1}).

In order to close the system of equations (\ref{field1}-\ref{field4}) we
have to give the equation of state for $p$ and specify $T$, $\xi $ and $\tau
$. As usual, we assume the following phenomenological (ad hoc) laws \cite
{Ma95}:
\begin{equation}
p=\omega \rho ,\quad \xi =k_{\gamma }\rho ^{\gamma },\quad T=D_{\delta }\rho
^{\delta },\quad \tau =\xi \rho ^{-1}=k_{\gamma }\rho ^{\gamma -1},
\label{csi4}
\end{equation}
where $0\leq \omega \leq 1$, and $k_{\gamma }\geq 0$, $D_{\delta }\geq 0$
are dimensional constants, $\gamma \geq 0$ and $\delta \geq 0$ ($\delta =%
\frac{\omega }{\omega +1}$ so that $0\leq \delta \leq 1/2$ for $0\leq \omega
\leq 1$) are numerical constants. Eqs. (\ref{csi4}) are standard in
cosmological models whereas the equation for $\tau $ is a simple procedure
to ensure that the speed of viscous pulses does not exceed the speed of
light. These are without sufficient thermodynamical motivation, but, in
absence of better alternatives, we use these equations and expect that they
will at least provide an indication of the range of possibilities. For the
temperature law, we take, $T=D_{\delta }\rho ^{\delta }$, which is the
simplest law guaranteeing positive heat capacity.

The growth of the total commoving entropy $\Sigma $ over a proper time
interval $\left( t_{0},t\right) $ is given by \cite{Ma95}:
\begin{equation}
\Sigma (t)-\Sigma \left( t_{0}\right) =-\frac{3}{k_{B}}\int_{t_{0}}^{t}\frac{%
\Pi Hf^{3}}{T}dt,  \label{M entropy}
\end{equation}
where $k_{B}$ is the Boltzmann's constant.

Therefore, with all these assumptions and taking into account the
conservation principle, i.e., $div(T_{i}^{j})=0$, the resulting field
equations are as follows:

\begin{align}
2\dot{H}+3H^{2}-2\frac{\dot{c}}{c}H& =-\frac{8\pi G}{c^{2}}\left( p+\Pi
\right) +\Lambda c^{2},  \label{nfield1} \\
3H^{2}& =\frac{8\pi G}{c^{2}}\rho +\Lambda c^{2},  \label{nfield2} \\
\dot{\rho}+3\left( \omega +1\right) \rho H& =-3H\Pi ,  \label{nfield3} \\
\frac{\dot{\Lambda}c^{4}}{8\pi G}+\rho \frac{\dot{G}}{G}-4\rho \frac{\dot{c}%
}{c}& =0,  \label{nfield4} \\
\dot{\Pi}+\frac{\Pi }{k_{\gamma }\rho ^{\gamma -1}}& =-3\rho H-\frac{1}{2}%
\Pi \left( 3H-W\frac{\dot{\rho}}{\rho }\right) ,  \label{nfield5}
\end{align}
where $W=\left( \frac{2\omega +1}{\omega +1}\right) $ is a numerical
constant.

\section{Lie method\label{LIE}}

The field equations (\ref{nfield1}-\ref{nfield5}) represent a system of odes
with 5 unknowns. In order to integrate them and to obtain a complete
solution for the proposed model we will need to make some simplifying
hypotheses (this is always a dangerous way) or to study if the system admits
symmetries. In this way, studying the possible symmetries that admit the
system, we will be able to determine the possible forms for which such
system is integrable.

In order to apply the standard Lie procedure (see for example \cite{Ovsi}-\cite
{Blumann}) we need to rewrite eq. (\ref{nfield1}-\ref{nfield5}) by a single
one and imposing that such ode (second order ode) admits a concrete symmetry
we will be able to determine the equation of the state for the viscous
parameter i.e. $\xi =k_{\gamma }\rho ^{\gamma }$ and the exact of the
``constants'' $G$ and $c$ for which the ode is completely integrable.

We start with the assumption $\Pi =\varkappa \rho $, with $\varkappa \in
\mathbb{R}^{-}$ (see \cite{Tony2} for a complete discussion of this
hypothesis).The bulk viscosity evolution equation can then be rewritten in
the alternative form
\begin{equation}
\delta \frac{\dot{\rho}}{\rho }+k_{\gamma }^{-1}\rho ^{1-\gamma }=-3\beta H,
\end{equation}
where $\beta =\left( \frac{1}{\varkappa }+\frac{1}{2}\right) $ and $\delta
=\left( 1-\frac{W}{2}\right) .$

Taking the derivative with respect to the time of this equation and with the
use of the equation
\begin{equation}
\dot{H}=\frac{\dot{c}}{c}H-4\pi \alpha \frac{G(t)}{c^{2}(t)}\rho ,
\end{equation}
obtained from the field equations (\ref{nfield1}) (where $\alpha =\left(
1+\omega +\varkappa \right) $). and taking into account eq. (\ref{nfield3})
\begin{equation}
H=-\frac{1}{3\alpha }\frac{\dot{\rho}}{\rho }
\end{equation}
we obtain the following second order differential equation describing the
time variation of the density of the cosmological fluid:
\begin{equation}
\ddot{\rho}=\frac{\dot{\rho}^{2}}{\rho }-As\rho ^{s}\dot{\rho}+B\frac{\dot{c}%
}{c}\dot{\rho}+D\frac{G}{c^{2}}\rho ^{2},  \label{neweq}
\end{equation}
where $A=\frac{k_{\gamma }^{-1}}{\delta },$ $s=(1-\gamma ),$ $B=\frac{\beta
}{\delta \alpha }$ and $D=\frac{12\pi \alpha \beta }{\delta }.$

We go next to apply all the standard procedure of Lie group analysis
to this equation (see \cite{Ibragimov} for details and notation). In this
way we find that the admissible symmetries for our ode being determinate by
the following system of pdes
\begin{align}
\rho \xi _{\rho \rho }+\xi _{\rho }& =0,  \label{ber1} \\
2\rho ^{2}\left( \frac{1}{2}\eta _{\rho \rho }-\xi _{t\rho }+\left( As\rho
^{s}-B\frac{\dot{c}}{c}\right) \xi _{\rho }\right) -\rho \eta _{\rho }+\eta
& =0,  \label{ber1_1} \\
2\eta _{t\rho }-c^{-2}\xi _{tt}-2\eta _{t}\rho ^{-1}+As^{2}\eta \rho
^{s-1}+\xi _{t}\left( sA\rho ^{s}-B\frac{\dot{c}}{c}\right) +\xi B\left( -%
\frac{\ddot{c}}{c}+\frac{\dot{c}^{2}}{c^{2}}\right) -3\rho ^{2}D\frac{G}{%
c^{2}}\xi _{\rho }& =0,  \label{ber3} \\
\eta _{tt}+\eta _{t}\left( sA\rho ^{s}-B\frac{\dot{c}}{c}\right) +D\frac{G}{%
c^{2}}\rho ^{2}\left( -2\xi _{t}+\eta _{\rho }-2\rho ^{-1}\eta \right) -D%
\frac{G}{c^{2}}\rho ^{2}\xi \left( \frac{\dot{G}}{G}-2\frac{\dot{c}}{c}%
\right) & =0,  \label{ber4}
\end{align}
Solving (\ref{ber1}-\ref{ber4}), we find that
\begin{equation}
\xi (\rho ,t)=at+b,\text{\ \ \ \ \ \ \ \ }\eta (\rho ,t)=-2a\rho ,\qquad
\Longleftrightarrow s=\frac{1}{2},
\end{equation}
with the constraints, from eq. (\ref{ber3})
\begin{equation}
\ddot{c}=\frac{\dot{c}^{2}}{c}-\frac{a}{at+2b}\dot{c},  \label{c-eq}
\end{equation}
and from eq. (\ref{ber4})
\begin{equation}
\frac{\dot{G}}{G}=2\frac{\dot{c}}{c},\qquad \Longrightarrow \frac{G}{c^{2}}%
=\kappa
\end{equation}
being $a$ and $b$ numerical constants and we will assume that $\kappa >0$.

In order to solve (\ref{c-eq}), we consider the following cases.

\begin{enumerate}
\item  Case I. Taking $a=0,b\neq 0,$ we get
\begin{equation}
c^{\prime \prime }=\frac{c^{\prime 2}}{c},\qquad \Longrightarrow \qquad
c(t)=K_{2}e^{K_{1}t},  \label{c_1}
\end{equation}
where the $\left( K_{i}\right) _{i=1}^{2}$ are integration constants.
Therefore the equation (\ref{neweq}) yields:
\begin{equation}
\ddot{\rho}=\frac{\dot{\rho}^{2}}{\rho }-\frac{A}{2}\sqrt{\rho }\dot{\rho}%
+BK_{1}\dot{\rho}+D\kappa \rho ^{2},  \label{eq1}
\end{equation}
the solution obtained through invariants is:
\begin{equation}
\frac{dt}{\xi }=\frac{d\rho }{\eta }\Longrightarrow \rho \thickapprox const.
\end{equation}
this solution seems not to be physical.

\item  Case II. Taking $b=0,a\neq 0,$ we get that the infinitesimal $X$ is $%
X=t\partial _{t}-2\rho \partial _{\rho \text{ }}$which is precisely the
generator of the scaling symmetries. Therefore the solution will be the same
than the obtained one with the dimensional method.

With these values of $a$ and $b$ equation (\ref{c-eq}) yields
\begin{equation}
c^{\prime \prime }=\frac{c^{\prime 2}}{c}-\frac{c^{\prime }}{t},\qquad
\Longrightarrow \qquad c(t)=K_{2}t^{K_{1}},  \label{c_2}
\end{equation}
as we expected, $c(t)$ follows a power-law solution. Therefore equation (\ref
{neweq}) yields:
\begin{equation}
\ddot{\rho}=\frac{\dot{\rho}^{2}}{\rho }-\frac{A}{2}\sqrt{\rho }\dot{\rho}+B%
\frac{K_{1}}{t}\dot{\rho}+D\kappa \rho ^{2},  \label{eq2}
\end{equation}
the solution obtained through invariants is:
\begin{equation}
\frac{dt}{\xi }=\frac{d\rho }{\eta }\Longrightarrow \rho \thickapprox
t^{-2},\qquad \rho =\rho _{0}t^{-2},  \label{sol_nl-c_2}
\end{equation}
where $\rho _{0}$ is a numerical constants that must verifies eq. (\ref{eq2}%
) in such a way that
\begin{equation}
\rho _{0}=\frac{1}{2}\left( \frac{A\left( A+\sqrt{A^{2}+8D\kappa \left(
1+BK_{1}\right) }\right) +4D\kappa \left( 1+BK_{1}\right) }{D^{2}\kappa ^{2}}%
\right) ,  \label{main}
\end{equation}
now, since $\rho _{0}$ must be a positive numerical constant, in a first
approach, and taking into account the definition of the constants $A,B$ and $%
D$ we see that $\rho _{0}>0$ iff $\left( 1+BK_{1}\right) <0,$ which imply
that $K_{1}>0$ since $A>0,B<0,\kappa >0$ and $D<0$ (since $\varkappa <0).$
Therefore $K_{1}$ must be a positive numerical constant.

Once again if one insists in solving equation (\ref{eq2}) through DA
(applying the Pi-theorem) it is found that \ with respect to the dimensional
base $\frak{B}=\left\{ \rho ,T\right\} $ each quantity has the following
dimensional equation $\left[ \rho \right] =\rho ,\left[ t\right] =T$ and $%
\left[ A\right] =\rho ^{-1}T^{-2}.$ Therefore, \ we find in \ a trivial way
that:
\begin{equation}
\begin{array}{r|rrr}
& \rho  & A & t \\ \hline
\rho  & 1 & -1 & 0 \\
T & 0 & -2 & 1
\end{array}
\Longrightarrow \rho \thickapprox \frac{1}{At^{2}}.
\end{equation}

\item  Case III. Taking $a,b\neq 0,$ we get
\begin{equation}
c^{\prime \prime }=\frac{c^{\prime 2}}{c}-\frac{ac^{\prime }}{at+b},\qquad
\Longrightarrow \qquad c(t)=K_{2}\left( at+b\right) ^{\frac{K_{1}}{a}},
\label{c_3}
\end{equation}
hence equation (\ref{neweq}) yields:
\begin{equation}
\ddot{\rho}=\frac{\dot{\rho}^{2}}{\rho }-\frac{A}{2}\sqrt{\rho }\dot{\rho}+B%
\frac{K_{1}}{at+b}\dot{\rho}+D\kappa \rho ^{2},  \label{eq3}
\end{equation}
the solution obtained through invariants is:
\begin{equation}
\frac{dt}{\xi }=\frac{d\rho }{\eta }\Longrightarrow \rho \thickapprox \left(
at+b\right) ^{-2/a},  \label{sol_nl-c_3}
\end{equation}

As we can see case II is a particular situation of case III.
\end{enumerate}

Through the symmetry analysis of eq. (\ref{neweq}) we have been able of
determining the equation of state that must follow the viscous parameter $%
\xi ,$ obtaining that $\gamma =1/2,$ as well as, that under the scaling
symmetry $c$ and $G$ must follow a power law solution in such a way that
they must verify the relationship $G/c^{2}=\kappa .$

\section{A Complete Solution}

Once we know the behaviour of the quantities $\rho ,G$ and $c$, in this
section, we are going to determine the rest of the quantities i.e. the scale
factor and mainly the behaviour of the entropy with the hope that this
quantity help us to determine the possible values of the constant $K_{1}.$
For this purpose we begin considering eq. (\ref{nfield3})
\begin{equation}
\dot{\rho}+3\left( \omega +1+\varkappa \right) \rho H=0,
\end{equation}
which trivially leads us to the well known relationship between the energy
density $\rho $ and the scale factor $f$
\begin{equation}
\rho =A_{\omega }f^{-3\left( \omega +1+\varkappa \right) }\text{ \ \ \ or \
\ \ \ \ }\rho =A_{\omega }f^{-3\alpha },  \label{INT conservation}
\end{equation}
where $\alpha =\left( \omega +1+\varkappa \right) .$ Now, taking into
account the eq. (\ref{nfield5})\ and simplifying it, we obtain,
\begin{equation}
\left( 1-\frac{W}{2}-\frac{1}{\varkappa \alpha }-\frac{1}{2\alpha }\right)
\dot{\rho}=-\frac{\sqrt[3]{\rho }}{k_{\gamma }},
\end{equation}
thereby obtaining $\rho =\rho (t).$ On simplifying further, we obtain,
\begin{equation}
\frac{\dot{\rho}}{\sqrt[3]{\rho }}=-\frac{1}{Kk_{\gamma }}\text{ \ \ \ }%
\Longrightarrow \text{ \ \ \ }\rho =\rho _{0}t^{-2},  \label{INT density}
\end{equation}
where
\begin{equation}
K=\left( 1-\frac{W}{2}-\frac{1}{\varkappa \alpha }-\frac{1}{2\alpha }\right)
,\text{ \ \ \ \ }\rho _{0}=\left( 2Kk_{\gamma }\right) ^{2}\text{.}
\end{equation}

From equation\ (\ref{INT conservation}) we obtain:
\begin{equation}
f=\left( \frac{A_{\omega }}{\rho _{0}}t^{2}\right) ^{1/3\alpha },\text{ \ \
i.e. \ \ }f\propto t^{\frac{2}{3\left( \omega +1+\varkappa \right) }}.
\label{INT sfactor}
\end{equation}

An important observational quantity is the deceleration parameter $q=\frac{d%
}{dt}\left( \frac{1}{H}\right) -1$. The sign of the deceleration parameter
indicates whether the model inflates or not. The positive sign of $q$
corresponds to ``standard'' decelerating models whereas the negative sign
indicates inflation. In our model, the deceleration parameter behaves as:
\begin{equation}
q=-1+\frac{3\alpha }{2}.
\end{equation}
In this way we find that for $\omega =1$, $\varkappa =-4/3$ is a critical
value since $q=0,$ therefore and under these considerations, $q<0$ if $%
\varkappa <-4/3$ and $q>0$ if $\varkappa >-4/3.$

We proceed with the calculation of the other physical quantities under the
condition $\gamma =1/2$:
\begin{align}
\xi & =k_{\gamma }\rho ^{\gamma }\propto k_{\gamma }\left( \rho
_{0}t^{-2}\right) ^{\gamma }\thickapprox k_{\gamma }^{2}t^{-1}, \\
T& =D_{\delta }\rho ^{\delta }=D_{\delta }\left( \rho _{0}t^{-2}\right)
^{\delta }\thickapprox t^{-2}\text{ \ \ \ with \ \ }\delta =\frac{\omega }{%
\omega +1}, \\
\tau & =\xi \rho ^{-1}=k_{\gamma }\left( \rho _{0}t^{-2}\right) ^{-1/2},%
\text{ i.e. }\tau =\left( 2K\right) ^{-1}t
\end{align}
We see from $\tau =\left( 2K\right) ^{-1}t$ that this result is in agreement
with the theoretical result obtained in \cite{Ma95}. For viscous expansion
to be non-thermalizing, we should have $\tau <t,$ or otherwise the basic
interaction rate for viscous effects should be sufficiently rapid to restore
the equilibrium as the fluid expands.

The commoving entropy is
\begin{equation}
\Sigma (t)=-\frac{2\varkappa }{\alpha }\Sigma _{0}\int^{t}x^{2\delta +\frac{2%
}{\alpha }-3}dx=\left( \omega +1\right) \Sigma _{0}t^{-\frac{\varkappa }{%
\alpha }}
\end{equation}
where $\Sigma _{0}=\frac{\rho _{0}^{1-\delta -1/\alpha }A_{\omega
}^{1/\alpha }}{D_{\delta }k_{B}}$ and we find that $\varkappa >-\left(
\omega +1\right) $ i.e. $\varkappa \in \left( -2,0\right] .$ We notice that
the parameter $\varkappa $ weakly perturbs the perfect fluid FRW Universe.
When $\varkappa =0$, the commoving entropy assumes a constant value and we
recover the perfect fluid case.

Finally, we will use equation (\ref{nfield4}) to obtain the behaviour of the
cosmological ``constant'' $\Lambda .$ Since we know the behaviour of the
constants $G$, $c$ and $\rho $ i.e.
\begin{equation}
G=\kappa c^{2},\qquad c(t)=K_{2}t^{K_{1}},\text{ \ \ \ and \ \ \ \ }\rho
=\rho _{0}t^{-2},  \label{mirian}
\end{equation}
then eq. (\ref{nfield4}) yields
\begin{equation}
\frac{\dot{\Lambda}c^{2}}{8\pi \kappa \rho }-2\frac{\dot{c}}{c}=0,
\label{ainoa}
\end{equation}
and substituting eq. (\ref{mirian}) into eq. (\ref{ainoa}) it is obtained
the following ODE
\begin{equation}
\dot{\Lambda}=16\pi \kappa \rho _{0}\frac{K_{1}}{K_{2}^{2}}%
t^{-2K_{1}-3}\qquad \Longrightarrow \qquad \Lambda =-8\pi \kappa \frac{\rho
_{0}}{K_{2}^{2}}\frac{K_{1}}{K_{1}+1}t^{-2\left( K_{1}+1\right) }
\label{cova}
\end{equation}
with $K_{1}>-1.$

We would like to emphasize that for $K_{1}>0$ then $\Lambda <0$ and for $%
K_{1}<0$ then $\Lambda >0$ but unfortunately we do not able to fix a better
value for the constant $K_{1}.$ Finally it is observed that
\begin{equation}
\Lambda \thickapprox \frac{1}{c^{2}t^{2}}.
\end{equation}
as it is expected in this self-similar solution.

As we have seen, with the followed way, we have not been able to determine a
better value for the constant $K_{1}$ than the obtained one in eq. (\ref
{main}). for this reason we try a new tentative which consists in
considering the effects of a $c-$variable into the scale factor in such a
way that when we calculate the entropy such effect may by reflected better
than in our previous way.

\subsection{A new tentative.}

In order to take into account the effects of a $c-$variable into the scale
factor we will determine it from eq. (\ref{nfield2}) since we already know
the behaviour of $G,c,\rho $ and $\Lambda .$ Therefore, after a
simplification eq. (\ref{nfield2}) yields$.$%
\begin{equation}
3H^{2}=8\pi \kappa \rho _{0}\left( 1-\frac{K_{1}}{\left( K_{1}+1\right) }%
\right) t^{-2},
\end{equation}
therefore
\begin{equation}
f=K_{f}t^{\upsilon }\qquad /\qquad \upsilon =\left( \frac{8\pi \kappa \rho
_{0}}{3}\frac{1}{K_{1}+1}\right) ^{1/2},
\end{equation}
finding again that $K_{1}>-1.$

In this way the entropy formula (\ref{M entropy}) yields:
\begin{equation}
\Sigma (t)=\Sigma _{0}t^{\frac{4\pi \kappa \rho _{0}}{K_{1}+1}+2\delta -2},
\end{equation}
where $\Sigma _{0}=-\varkappa \sqrt{6}\rho _{0}\frac{K_{f}^{3}}{k_{B}}\frac{%
\sqrt{\left( \pi \kappa \rho _{0}\left( K_{1}+1\right) \right) }}{2\pi
\kappa \rho _{0}+\left( K_{1}+1\right) \left( \delta -1\right) }.$ If we
make $2\pi \kappa \rho _{0}=1$ (a free hypothesis), $\Sigma (t)\thickapprox
\Sigma _{0}t^{2\left( \frac{1}{K_{1}+1}-\frac{1}{\omega +1}\right) }$ we
find therefore $K_{1}>-\omega ,$ but we are not able to find a better value
(or limit) for $K_{1}.$

Or taking into account the field eq. (\ref{nfield1})
\begin{equation}
2\dot{H}+3H^{2}-2\frac{K_{1}}{t}H=-8\pi \kappa \rho _{0}\left( \left( \omega
+\varkappa \right) +\frac{K_{1}}{K_{1}+1}\right) t^{-2}
\end{equation}
which is a Riccati ode and which particular solution (scaling solution) is:
\begin{equation}
H=\frac{a}{t}\qquad \Longrightarrow \qquad f=K_{f}t^{a}
\end{equation}
where
\begin{equation}
a=\frac{\left( K_{1}+1\right) ^{2}\pm \sqrt{\left( K_{1}+1\right)
^{4}-3A\left( \alpha \left( K_{1}+1\right) ^{2}+K_{1}^{2}+K_{1}\right) }}{%
3\left( K_{1}+1\right) }
\end{equation}
finding that $K_{1}>-1,$ with $\alpha =\left( \omega +\varkappa \right) $, $%
A=8\pi \kappa \rho _{0}$ and imposing the condition$.$%
\begin{equation}
\left( K_{1}+1\right) ^{4}-3A\left( \alpha \left( K_{1}+1\right)
^{2}+K_{1}^{2}+K_{1}\right) \geq 0
\end{equation}
(note that $\alpha $ may take negative values) therefore the entropy behaves
as:
\begin{equation}
\Sigma \thickapprox t^{-2+2\delta +\frac{\left( K_{1}+1\right) ^{2}+\sqrt{%
\left( K_{1}+1\right) ^{4}-3A\left( \alpha \left( K_{1}+1\right)
^{2}+K_{1}^{2}+K_{1}\right) }}{\left( K_{1}+1\right) }}
\end{equation}

As we can see following these ways we have only been able to determine that $%
K_{1}>-1$ or that
\begin{equation}
\psi (K_{1})>\left| \frac{2}{\omega +1}\right|
\end{equation}
but as $\psi (K_{1})$ depends on the factors like $A$ and $\alpha $ it is
very difficult to determine a good value for this constant.

\section{Conclusions.}

In this paper we have studied a flat FRW cosmological model filled with a
bulk viscous fluid and where the ``constants'' $G,c$ and $\Lambda $ are
considered as functions on time $t.$ In order to outline the field equations
we have taken into account the possible effects of a $c-$variable into the
curvature tensor. In this way we have been able to deduce the general
conservation principle from the field equations \ obtaining an equation that
coincides with the obtained one from the divergence of the right hand of the
field equation i.e. $div\left( \frac{G}{c^{4}}T_{i}^{j}+\delta
_{i}^{j}\Lambda \right) =0.$ Using the Lie group tactic we have been able to
determine the equation of state for the bulk viscous parameter, $\xi
=k_{\gamma }\rho ^{\gamma },$ as well as the exact form for the
``constants'' $G$ and $c$ which make integrable the field equations.

We have obtained that our model is self-similar i.e. admits a scaling
symmetry iff $\gamma =1/2$ and $c$ follows a power law solution i.e. $%
c=c_{0}t^{K_{1}},$ while $G$ and $c$ must verify the relationship $%
G/c^{2}=\kappa .$ Under these results we have calculated the behaviour of the
energy density finding, in a first approach, that the physical solution is
only possible if $K_{1}>0,$ that is to say, $G$ and $c$ are growing
functions on time $t.$ We have also calculated the rest of the main
quantities but under physical considerations we have only been able to
determine that $K_{1}>-1.$ The latter result does not contradict our previous
result $K_{1}>0.$

Therefore we conclude that $G$ and $c$ are growing functions while
$\Lambda $ is a negative decreasing function. \vspace{1cm}

 \textbf{Acknowledgement.} I would wish to express my gratitude to
Javier Aceves for his translation into English of this paper.

\end{document}